\documentclass[useAMS,usenatbib]{mn2e}
\usepackage{graphicx}
\usepackage{url}

\def\lsim{\,\lower2truept\hbox{${<\atop\hbox{\raise4truept\hbox{$\sim$}}}$}\,}
\def\gsim{\,\lower2truept\hbox{${> \atop\hbox{\raise4truept\hbox{$\sim$}}}$}\,}
\def\simlt{\mathrel{\rlap{\lower 3pt\hbox{$\sim$}}\raise 2.0pt\hbox{$<$}}}
\def\simgt{\mathrel{\rlap{\lower 3pt\hbox{$\sim$}} \raise2.0pt\hbox{$>$}}}

\title[Statistics of proto-clusters]
  {On the statistics of proto-cluster candidates detected in the {\it Planck} all-sky survey}
\author[M. Negrello et al.]
{M.~Negrello$^{1}$\thanks{NegrelloM@cardiff.ac.uk},
J.~Gonzalez-Nuevo$^{2}$,
G.~De Zotti$^{3}$,
M.~Bonato$^{3,4}$,
Z.-Y.~Cai$^{5, 6}$,
\newauthor
D. Clements$^{7}$,
L.~Danese$^{4}$,
H.~Dole$^{8,9}$,
J.~Greenslade$^{5}$,
A.~Lapi$^{4}$,
L.~Montier$^{10,11}$
\\
$^{1}$School of Physics and Astronomy, Cardiff University, The Parade,
Cardiff CF24 3AA, UK \\
$^{2}$Departamento de F\'{\i}sica, Universidad de Oviedo, Avda. Calvo Sotelo s/n, Oviedo, Spain \\
$^{3}$INAF, Osservatorio Astronomico di Padova, Vicolo Osservatorio 5, I-35122 Padova, Italy \\
$^{4}$SISSA, Via Bonomea 265, 34136 Trieste, Italy \\
$^{5}$CAS Key Laboratory for Research in Galaxies and Cosmology,
Department of Astronomy, \\ University of Science and Technology of
China, Hefei 230026, China \\
$^{6}$School of Astronomy and Space Science, University of Science and
Technology of China, Hefei 230026, China \\
$^{7}$Imperial College London, Blackett Laboratory, Prince Consort Road, London SW7 2AZ, UK \\
$^{8}$Institut d'Astrophysique Spatiale, CNRS (UMR 8617) Universit\'e Paris-Sud 11, B\^atiment 121, 91405 Orsay, France \\
$^{9}$Institut Universitaire de France, 103 bd Saint-Michel, 75005 Paris, France \\
$^{10}$Universit\'e de Toulouse, UPS-OMP, IRAP, 31028 Toulouse Cedex 4, France \\
$^{11}$CNRS, IRAP, 9 Av. colonel Roche, BP 44346, 31028 Toulouse Cedex 4, France
}

\pagerange{\pageref{firstpage}--\pageref{lastpage}} \pubyear{2002}

\def\LaTeX{L\kern-.36em\raise.3ex\hbox{a}\kern-.15em
    T\kern-.1667em\lower.7ex\hbox{E}\kern-.125emX}

\begin{document}

\label{firstpage}

\maketitle

\begin{abstract}
Observational investigations of the abundance of massive precursors of
local galaxy clusters (``proto-clusters'') allow us
to test the growth of density perturbations, to constrain cosmological
parameters that control it, to test the theory of non-linear collapse
and how the galaxy formation takes place in dense environments.
The \textit{Planck} collaboration has recently published a catalogue
of $\gsim$\,2000
{\it cold} extra-galactic sub-millimeter sources, i.e. with colours
indicative of $z\gsim2$, almost all of which
appear to be over-densities of star-forming galaxies.
They are thus considered as proto-cluster candidates. Their number densities (or their flux densities) are far in
excess of expectations
from the standard scenario for the evolution of large-scale
structure.
Simulations based on a physically motivated galaxy evolution model
show that essentially all {\it cold}  peaks brighter than
$S_{545\rm GHz}=500\, $mJy  found in {\it Planck} maps after having
removed the Galactic dust emission can be interpreted as positive Poisson fluctuations of the number of high-$z$ dusty proto-clusters
within the same \textit{Planck} beam, rather then being individual clumps of physically bound galaxies.
This conclusion does not change if an empirical fit to the luminosity
  function of dusty galaxies is used instead of the physical model.
The simulations accurately reproduce
the statistic of the {\it Planck} detections and yield distributions
of sizes and ellipticities in qualitative agreement with observations.
The redshift distribution of the brightest proto-clusters contributing
to the {\it cold} peaks has a broad maximum at $1.5 \le z  \le 3$.
Therefore follow-up of \textit{Planck} proto-cluster candidates will  provide key information on the high-$z$ evolution of large scale structure.
\end{abstract}

\begin{keywords}
galaxies: clusters: general -- galaxies: evolution -- large-scale structure of Universe -- submillimetre: galaxies
\end{keywords}

\section{Introduction}\label{sec:intro}

Observations of the early evolution of large scale structure are
important tools for testing galaxy formation theories,
investigating galaxy evolution in dense environments and getting
independent constraints on fundamental cosmological
parameters \citep[e.g.,][]{HarrisonColes2012,Casey2016}. This has
motivated several observational efforts to identify
high-redshift proto-clusters of galaxies \citep[for a review see][]{Overzier2016}
defined as high-$z$ galaxy over-densities that will evolve into the local galaxy clusters.

The detection of these objects, however, proved to be
challenging. Above $z\simeq 1.5$
the searches via the X-ray emission or the Sunyaev-Zeldovich (SZ)
effect as well as the optical  detection of the galaxy red sequence are inefficient, not only because of the
difficulty of high-$z$  observations but even more because these methods are appropriate to
detect evolved clusters with mature galaxy  populations or the signature of a hot inter-cluster medium (ICM). But
in high-$z$ proto-clusters the ICM may not have yet reached the high
temperatures  necessary for X-ray or SZ detections,
and member galaxies may be still in the active star-formation phase.

Indeed, (sub-)millimeter surveys with \textit{Herschel}, SCUBA-2 or
APEX LABOCA proved to be capable of detecting
candidate high-$z$ proto-clusters of star-bursting galaxies
\citep{Valtchanov2013,Dannerbauer2014,Rigby2014,Ma2015,Casey2015,Clements2016}.
These objects are however rare and therefore very large area surveys
are necessary to get a good statistics. As first pointed out by
\citet{Negrello2005} low-resolution sub-mm surveys, such as those
carried out by the \textit{Planck} satellite,
are very well suited for detecting proto-clusters because they integrate over the emission of \textit{all} star-forming member galaxies, not only of those above some detection limit.

The \citet[][PCXXXIX
hereafter]{PlanckCollaborationXXXIX2015} provided a catalog of 2151 cold sub-mm sources with spectral energy
distributions peaking between 353 and 857 GHz at $5'$ resolution,
detected by \textit{Planck} in the cleanest 26\% of the sky, with flux
density at 545 GHz above 500\,mJy. This is referred to as the
{\it Planck} high-$z$, or PH$z$, sample.
The vast majority of the
sources in the sample appear as overdensities of dusty star-forming galaxies,
having colours consistent with $z\ge 2$ and were considered as
proto-cluster candidates.

\citet{PlanckCollaborationXXVII2015} used the Spectral and Photometric Imaging Receiver (SPIRE) on the \textit{Herschel} satellite to perform follow-up observations of 234 {\it cold} \textit{Planck}
sources, satisfying the criterion of having their rest-frame
far-IR peak redshifted to the frequency range $353-857$\,GHz, as
expected for galaxies in the redshift range $z=2$--4. Usable results
were obtained for 228 of them, including 83 of the 203 sources within
the area covered by the PH$z$ sample. About 94\% of the SPIRE sources
associated to the 228 \textit{Planck} objects were found to be
consistent with being galaxy overdensities; their (photometric) redshift
distribution peaks at $z\simeq 2$ for a typical dust temperature of
35\,K. Seven sources (about 3\%) are candidate lensed systems; all of
them have spectroscopic redshifts and are at $z>2.2$
\citep{Canameras2015}.

As pointed out by PCXXXIX the nature of the high-$z$ galaxy
overdensities is still uncertain. Given the large uncertainties in the
photometric redshift estimates, it is impossible, at present, to
establish whether galaxies making up the overdensities are physically
related or are just part of random fluctuations in the galaxy number
density within the \textit{Planck} resolution element.

PCXXXIX found that the distribution of the total (8--$1000\,\mu$m)
infrared (IR) luminosities of overdensities peaks around $2\times
10^{14}\,L_\odot$ for their reference dust temperature of 35\,K. Using
the calibration by \citet{KennicuttEvans2012}, the corresponding star
formation rate (SFR) is $\simeq 3\times 10^4\,M_\odot\,\hbox{yr}^{-1}$. The associated halo mass, $M_{\rm h}$, can be estimated using the SFR--$M_{\rm h}$ relation at $z=2$ derived by \citet{Aversa2015} exploiting the abundance matching technique. For mean SFRs in the range 30--1000\,$M_\odot\,\hbox{yr}^{-1}$ \citep[the typical SFR of $z\simeq 2$ star-forming galaxies is $\simeq 300\,M_\odot\,\hbox{yr}^{-1}$, see Fig.~9 of][]{Cai2014} the sum of halo masses of star-forming galaxies is $M_{\rm h,sf}\simeq 5\times 10^{14}\,M_\odot$, independently of the SFR. If the lifetime of the star-forming phase is $t_{\rm sf}$, the \textit{total} halo mass is $M_{\rm h}\simeq M_{\rm h,sf}({t_{\rm u}/t_{\rm sf}})$, where $t_{\rm u}\simeq 3.3\,$Gyr is the age of the universe at $z\simeq 2$. Since $t_{\rm sf}$ is likely $< 1\,$Gyr, $M_{\rm h}$ is probably several times larger than $M_{\rm h,sf}$.  The redshift-dependent halo mass function by \cite{ST99} gives a surface density of halos with mass larger than $M_{\rm h,sf}$ at $z>2$ of $\sim5\times10^{-5}$\,deg$^{-2}$ (and much lower than that for $M_{\rm h}$, since the density sinks down exponentially in this mass range) to be compared with the surface density of
\textit{Planck} overdensities of $0.21\,\hbox{deg}^{-2}$. This already
highlights a problem with the interpretation of overdensities as
proto-clusters.

To look for a plausible explanation of the results by
PCXXXIX we have updated the study by
\citet{Negrello2005} who worked out a formalism to derive the IR
luminosity functions and number counts of proto-clusters of dusty
galaxies. In the last decade the amount of data on the cosmological
evolution of the IR luminosity function and on the clustering
properties of dusty galaxies has grown enormously, allowing us to put
our estimates on a much more solid basis than was possible in 2005. We
will adopt as our reference the latest \cite{Cai2013} version of the
self-regulated galaxy evolution model developed by \cite{Granato2004} and \cite{Lapi2006,Lapi2011}.
However we checked that our results and conclusions do not significantly change
if we adopt an empirical fit the luminosity function of dusty
galaxies \citep[e.g.][]{Mancuso2016}.

The paper is organized as follow. In Section\,\ref{sec:formalism} we
review and update the formalism introduced by \citet{Negrello2005}, while in
Section\,\ref{sec:galaxy_model} we briefly describe the model adopted
for the luminosity function and the clustering properties of dusty
galaxies. Predictions for the number counts of proto-clusters and
comparison with data are presented in Section\,\ref{sec:results}. In
Section\,\ref{sec:simulations} we investigate the effect on the estimated number
counts of Poisson fluctuations of the number of proto-clusters within the \textit{Planck} beam. Section\,\ref{sec:conclusions} summarizes and discusses our main conclusions.

Throughout the paper we assume a flat cold dark matter cosmology with
$\Omega_{0,m}=0.315$, $\Omega_{0,b}=0.044$, $h=0.67$ and
$\sigma_{8}=0.81$, with an index $n=1$ for the power spectrum of primordial density fluctuations, consistent with the \textit{Planck} results \citep{PlanckCollaborationXIII2016}.

\section{Formalism}\label{sec:formalism}

The mean luminosity of
a clump of {\it clustered} galaxies, or of a {\it proto-cluster}, at
redshift $z$ is\footnote{At variance with \cite{Negrello2005} we have now
  dropped from the equations all the Poisson terms, as we are
   specifically interested in {\it physically related} galaxies.} \citep{Peebles1980}
\begin{equation}\label{eq:meanL_clump}
\bar{L}_{\rm cl}(z) = \int_{\mathcal L}
\frac{dN(L^{\prime},z)}{dL^{\prime}dV} L^{\prime} dL^{\prime} \cdot
\int_{V_{\rm cl}}\xi(r,z) dV,
\end{equation}
where $dN(L,z)/dLdV$ is the comoving luminosity function of galaxies
and $\xi(r,z)$ is their spatial correlation function; $V_{\rm cl}$ is
the volume of the proto-cluster (the choice of the proto-cluster
radius will be discussed in Section\,\ref{sec:galaxy_model}), while
$r$ is the comoving radial distance to the volume element $dV$.
The variance of the clump luminosity is
\begin{eqnarray}\label{eq:sigma2_clump}
\sigma^{2}_{L_{\rm cl}}(z) & = & \int_{\mathcal L}
                             \frac{dN(L^{\prime},z)}{dL^{\prime}dV}
                             L^{\prime 2} dL^{\prime} \cdot
                             \int_{V_{\rm cl}}\xi(r,z)dV
\nonumber \\
& + & \left[ \int_{\mathcal L} \frac{dN(L^{\prime},z)}{dL^{\prime}dV}
L^{\prime} dL^{\prime} \right]^{2} \cdot  \int\int_{V_{\rm cl}}[\zeta(r_{1},r_{2},z)  \nonumber \\
& +&\xi(r_{12},z) - \xi(r_{1},z)\xi(r_{2},z)]dV_{1}dV_{2}
\end{eqnarray}
where $r_{12}$ is the comoving distance between the volume elements
$dV_{1}$ and $dV_{2}$, and $\zeta$ is the reduced part of the three-point spatial correlation function for which we assume the standard hierarchical formula:
\begin{equation}\label{eq:zeta}
\zeta(r_{1},r_{2}) = Q(z)\left[  \xi(r_{1})\xi(r_{2}) + \xi(r_{1})\xi(r_{12}) + \xi(r_{12})\xi(r_{2}) \right].
\end{equation}
The variance of clump luminosities [eq.~(\ref{eq:sigma2_clump})] is the sum of the contributions from fluctuations in the total
luminosity and in the number of member sources. ). \\
Using Eq.\,(\ref{eq:zeta}) we rewrite Eq.\,(\ref{eq:sigma2_clump}) as
\begin{eqnarray}\label{eq:sigma2}
\sigma^{2}_{cl}(z) & = &  \int_{\mathcal L} \frac{dN(L^{\prime},z)}{dL^{\prime}dV}
L^{\prime 2} dL^{\prime} \cdot \int_{V}\xi(r,z) dV \nonumber \\
& + & \left[ \int_{\mathcal L} \frac{dN(L^{\prime},z)}{dL^{\prime}dV}
L^{\prime} dL^{\prime} \right]^{2} \cdot
\nonumber \\
&~& [Q(z) - 1] \cdot \left(  \int_{V}\xi(r,z) dV \right)^{2}  \nonumber \\
& + & \left[ \int_{\mathcal L} \frac{dN(L^{\prime},z)}{dL^{\prime}dV}
L^{\prime} dL^{\prime} \right]^{2} \cdot \nonumber \\
&~& 2Q(z) \cdot \int\int_{V}\xi(r_{12},z)\xi(r_{1}) dV_{1}dV_{2}  \nonumber \\
& + & \left[ \int_{\mathcal L} \frac{dN(L^{\prime},z)}{dL^{\prime}dV}
L^{\prime} dL^{\prime} \right]^{2} \cdot \nonumber \\
&~& \int\int_{V}\xi(r_{12},z) dV_{1}dV_{2}.
\end{eqnarray}
The statistics of the matter-density distribution found in N-body
simulations is succesfully reproduced by a log-normal function
\citep{ColesJones1991,Kofman1994,TaylorWatts2000,Kayo2001,Taruya2003}, not only in weakly non-linear regimes, but also up to
density contrasts $\delta \approx 100$. If light is a (biased) tracer
of mass, fluctuations in the luminosity density should reflect the
statistics of the matter-density field. Therefore we adopt, following
N05, a log-normal shape for the distribution of $L_{\rm cl}$. This
function is completely determined by its first (mean) and second
(variance) moments:  
\begin{eqnarray}\label{eq:LF_clumps}
\frac{dN(L_{\rm cl},z)}{dL_{\rm cl}dV} = \frac{\exp[-\frac{1}{2}[\ln{L_{\rm cl}
    - \mu(z)}]^{2}/\sigma^{2}(z)]}{\sqrt{2\pi\sigma^{2}(z)} L_{\rm cl}},
\end{eqnarray}
where
\begin{eqnarray}
\mu(z) & = & \ln{ \left[ \frac{\bar{L}^{2}_{\rm cl}(z)}{\sqrt{\sigma^{2}_{L_{\rm cl}}(z)
      + \bar{L}^{2}_{\rm cl}(z)}} \right] }, \\
\sigma^{2}(z) & = & \ln{ \left[ \frac{\sigma^{2}_{L_{\rm cl}}(z)}{\bar{L}^{2}_{\rm cl}(z)} + 1 \right] }.
\end{eqnarray}
The function is normalized by requiring the conservation of the luminosity density
\begin{eqnarray}\label{eq:lum_dens}
\int_{\mathcal L_{\rm cl}}\frac{dN(L^{\prime}_{\rm cl},z)}{dL^{\prime}_{\rm cl}dV}L_{\rm cl}^{\prime}dL_{\rm cl}^{\prime}
= \int_{\mathcal L}\frac{dN(L^{\prime},z)}{dL^{\prime}dV}L^{\prime}dL^{\prime}.
\end{eqnarray}

\begin{figure}
\vspace{-3.0cm}
\hspace{-1.0cm}
\makebox[\textwidth][c]{
\includegraphics[width=0.95\textwidth]{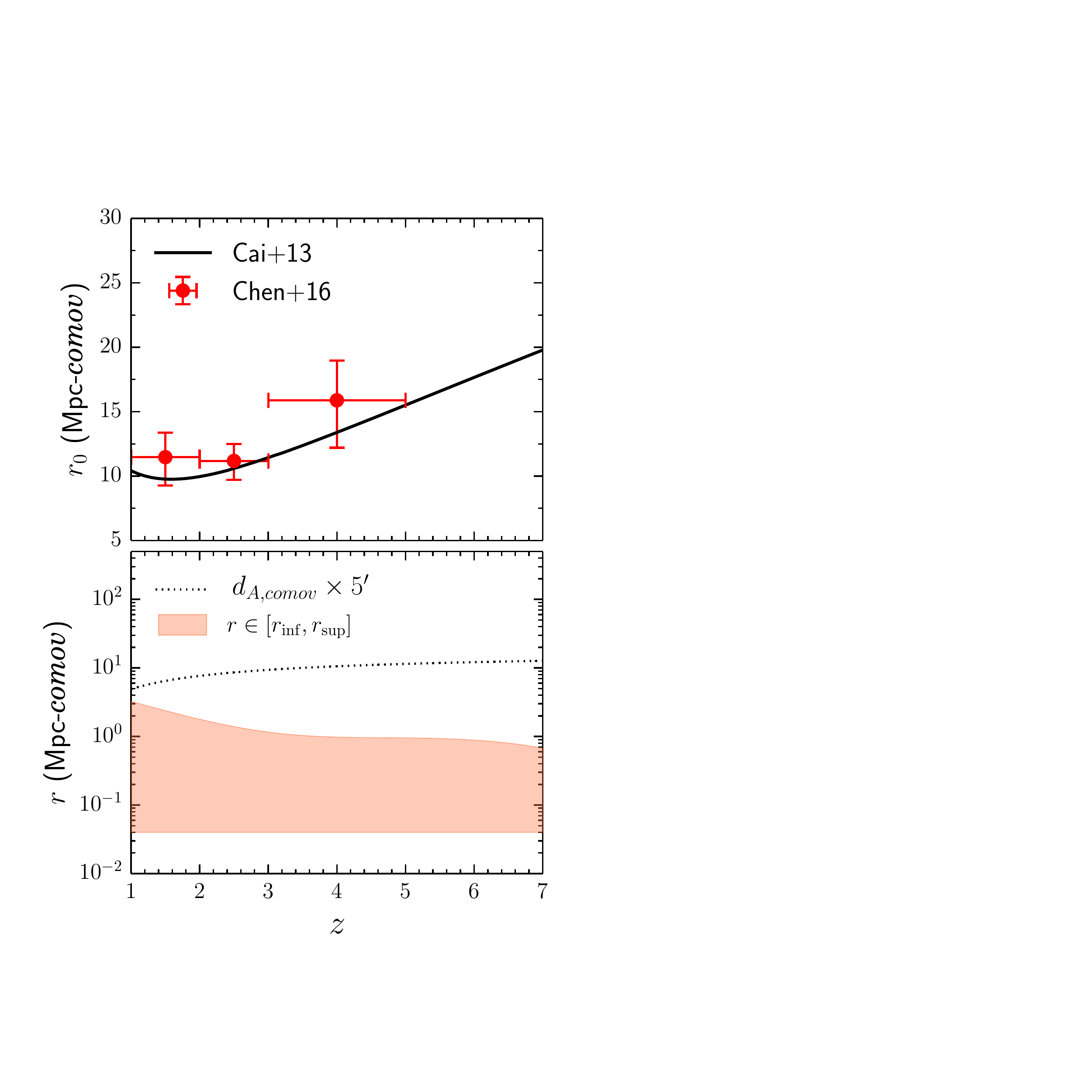}}
\vspace{-2.8cm}
\caption{{\it Upper panel}: spatial correlation length of proto-spheroids
  estimated by Cai et al. (2013; solid line) compared to the observational determination by Chen
  et al. (2016) from a sample of near-infrared selected sub-millimeter
galaxies (red dots). {\it Lower panel}: lower and upper limits on the
value of $r$ adopted for integration in eqs.\,(\ref{eq:meanL_clump}) and (\ref{eq:sigma2}). The dotted curve shows, for comparison, the
comoving scale corresponding to an angular separation of
5$^{\prime}$, approximately the FWHM of the {\it   Planck} beam at 545\,GHz.}
\label{fig:scales}
\end{figure}

\begin{figure*}
\vspace{0.0cm}
\hspace{0.0cm}
\makebox[\textwidth][c]{
\includegraphics[width=0.95\textwidth]{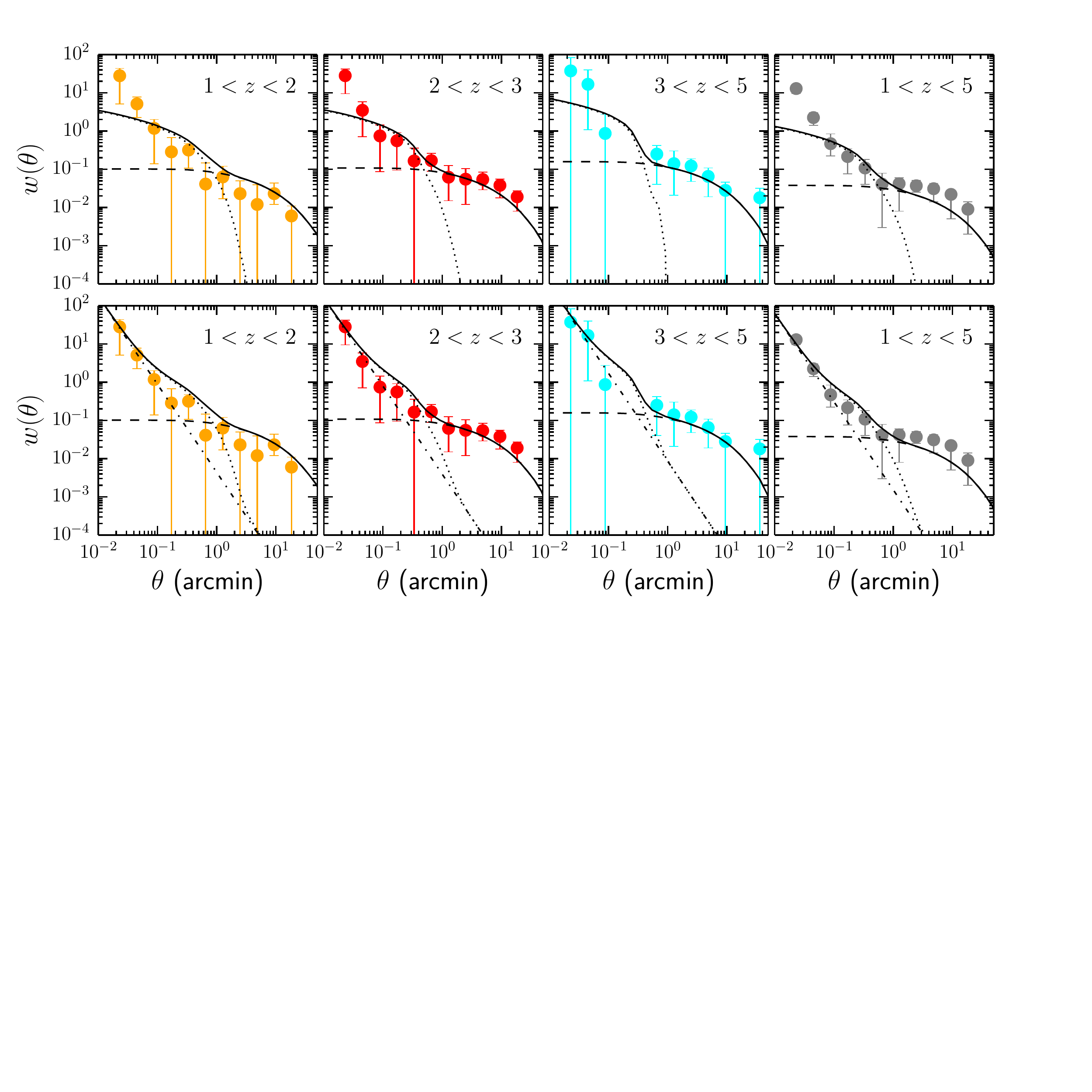}}
\vspace{-8.0cm}
\caption{Angular correlation function of faint submillimeter galaxies
  measured by Chen et al. (2016; data points) in different photometric
  redshift bins. {\it Top panels}: comparison with the theoretical
  angular correlation function derived from the model of Cai et
  al. (2013; solid black curve) without any fine tuning of the clustering
  parameters. The 1- and 2-halo terms are shown by the dotted and the
  dashed curves, respectively. {\it Bottom panels}: same as above but with
  an additional contribution to the 1-halo term, modelled as a power-law
  ($\xi(r)\propto r^{-\gamma}$, with $\gamma=3.3$, dot-dashed curve; see text).}
\label{fig:wtheta}
\end{figure*}

\begin{figure}
\hspace{-5.0cm}
\makebox[\textwidth][c]{
\includegraphics[width=0.45\textwidth]{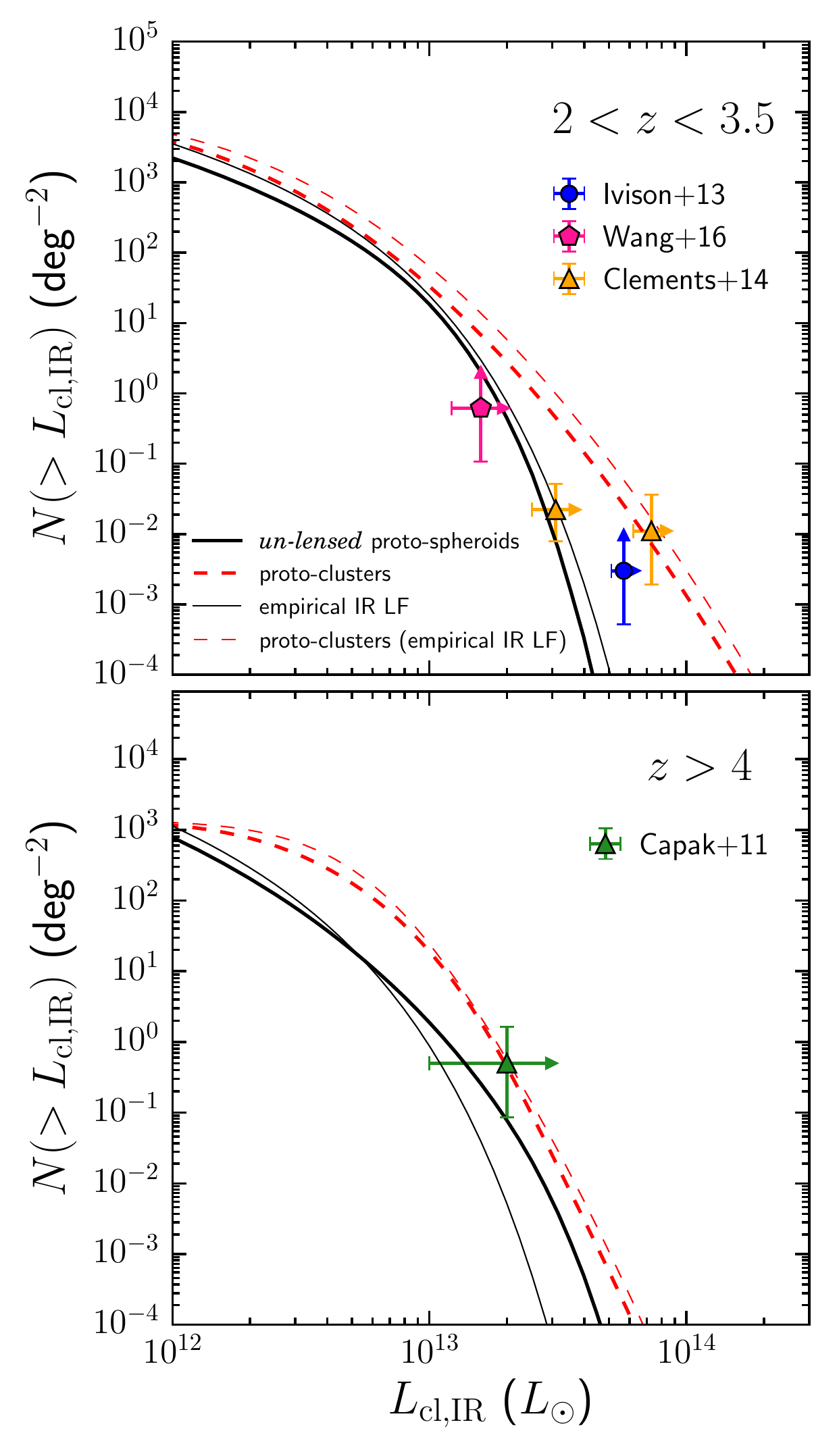}}
\vspace{-0.5cm}
\caption{Cumulative infrared luminosity function of proto-clusters with
  redshifts in the range $2<z<3.5$ (upper panel) and with $z>4$ (lower panel). The
  predictions, based on the Cai et al. model are shown by the thick dashed red curve while, for comparison, the abundance of
un-lensed protospheroids is shown by the thick solid black curve. The
excess density of proto-clusters compared to unlensed protospheroids
is compensated by a deficit at faint luminosities (not shown): the
luminosity density is conserved [eq.~(\protect\ref{eq:lum_dens})].
The lighter curves are predictions based on the infrared
  luminosity function of Mancuso et al. (2016) for dusty
  galaxies (solid back curve) and proto-clusters (dashed red curve).
The data
points mark the constraints on the abundance of
  dusty proto-clusters set by the discoveries of such objects reported
  in Capak et al. (2011), Ivison et al. (2013), Clements et al. (2014)
  and Wang et al. (2016). The luminosities of detected proto-clusters are the sum of those of member galaxies above the detection limits. Hence, they do not include the contribution of fainter member galaxies, whose summed emission may be dominant (see text).  }
\label{fig:NgtLirclump}
\end{figure}

\section{Adopted model for dusty galaxies}\label{sec:galaxy_model}

We compute the clump total-infrared ($8-1000\,\mu$m) luminosity
function using the redshift--dependent IR luminosity function of dusty
galaxies provided by the \citet[][C13 hereafter]{Cai2013} model which accurately
reproduces the observational determinations, available up to $z\simeq
4$ \citep[see Fig.~1 of][]{Bonato2014}, as well as a broad variety of
other relevant data\footnote{A tabulation of the IR luminosity
  function at several redshifts as well as figures showing fits to
  other relevant data sets are available at
  \url{http://people.sissa.it/~zcai/galaxy_agn/} or
  \url{http://staff.ustc.edu.cn/~zcai/galaxy_agn/index.html}.}.
In order to assess the robustness of our results
  we also work out predictions based on a parametric fit to the
  infrared luminosity function of dusty galaxies. Specifically we
  use the results by \cite{Mancuso2016} who preformed a fit to the global, redshift-dependent, 
  SFR function of galaxies measured at UV and far-IR
  wavelengths by the {\it Hubble} and the {\it Herschel} space
  telescopes. As we are interested in the dust-obscured phase of the
  formation of galaxy clusters at $z\gsim1$, we just need the part of
  the global SFR function above $SFR\gsim50\,$M$_{\odot}\,$yr$^{-1}$ [corresponding to
  luminosities $L_{\rm SFR}\gsim few\times10^{11}\,$L$_{\odot}$
  \citep{KennicuttEvans2012}], where, generally, the star formation is heavily obscured by dust. In practise we identify the value of the SFR at which the global SFR function is a factor of 10
 higher than the SFR function derived by Mancuso et al. from UV
 (dust-corrected) data alone and apply an exponential cut-off below
 that value. The cut-off correponds to $SFR\sim60\,$M$_{\odot}$\,yr$^{-1}$ to $120$\,M$_{\odot}$\,yr$^{-1}$ from $z=2$ to
 $z=5$, or to infrared luminosities $L_{\rm
   IR}\sim4\times10^{11}\,$L$_{\odot}$ to
 $\sim8\times10^{11}\,$L$_{\odot}$, respectively. \\

The two-point correlation function of dusty galaxies was computed
using the Halo Occupation Distribution (HOD) formalism.  The
power spectrum of the galaxy distribution is parameterized as the sum
of two terms: a 1-halo term, dominating on small scales, that accounts for the
distribution of galaxies within the same halo, and a 2-halo term,
that describes the correlations among galaxies residing on
different halos and, therefore, dominates on larger scales. The HOD
provides a statistical description of how
galaxies populate dark matter halos and we model it using the
central-satellite formalism \citep[see, e.g.,][]{Zheng2005}. This
assumes that the first galaxy to be hosted by a halo lies at its
center, while the remaining galaxies
are distributed in proportion to the halo mass profile and are
classified as satellites. The mean halo
occupation function of satellite galaxies is parameterized as:
$\langle{N_{\rm sat}}\rangle \propto (M_{\rm vir}/M_{\rm
  sat})^{\alpha_{\rm sat}}$, where $M_{\rm vir}$ is the halo mass and
the power-law index $\alpha_{\rm sat}$ is a free parameter. As for 
the 2-halo term, the key parameter, determining the amplitude of the effective bias function
$b_{\rm eff}(z)$, is the minimum halo mass, $M_{\rm vir, min}$.

We have adopted the HOD parameter values derived by C13 by
fitting the power spectra of the Cosmic Infrared Background (CIB)
measured by \textit{Planck} and \textit{Herschel}. The spatial
correlation function, $\xi(r,z)$, was then computed as the Fourier
anti-transform of the 3D power spectrum. We refer the reader to
\citet{Xia12}  for all the relevant details on
the formalism used. The clustering radius,
or correlation length,
$r_0$, defined by $\xi(r_{0},z)=1$, is found to
be, in comoving units, $r_0\simeq 10-20\,$Mpc in the redshift range
$z\sim1-7$. It is shown by the solid line in the upper panel of
Fig.~\ref{fig:scales}.

Recently, \cite{Chen2016b} have measured the
correlation length of a sample of $\sim3000$ sub-millimetre galaxies, with redshifts $z\sim1-5$
and star formation rates $\gsim60-100$M$_{\odot}\,$yr$^{-1}$, identified using a new
colour selection technique, which combines three optical-near-infrared
colours \citep{Chen2016a}. Their estimates of $r_0$ in 3 redshift intervals ($1< z <2$, $2< z <3$ and $3< z <5$) are shown in the same
figure and are in good agreement with the results of C13.

In Fig.\,\ref{fig:wtheta} we show the angular correlation function measured by \cite{Chen2016b} in the 3 redshift slices and for the full sample ($1< z <5$). The prediction of the C13 model (computed using the redshift distribution of the Chen et
al. sample), without any tuning of the HOD parameters, is also shown for comparison (dashed
curve for the 2-halo term; dotted curve for the 1-halo term). The
agreement is remarkably good on angular scales
$\theta\gsim0.1^{\prime}$; however, on the smallest scales the model significantly
underestimate the clustering of sub-mm galaxies. This is not a
surprise considering that the constraints derived by C13 were obtained from measurements performed at larger angular scales than those probed by Chen et al. It is beyond
the scope of this paper to perform a detailed  analysis of the signal measured by
Chen et al. within the HOD formalism. Therefore, in order to achieve a good agreement with the data we add an extra contribution to
the 1-halo term below $\theta=$0.2$^{\prime}$, that we model as a power-law,
$\xi(r)\propto r^{-\gamma}$.  The fit to the data gives
$\gamma=3.3\pm0.2$. The extra term is shown by the dot-dashed curve in
the lower panels of Fig.\,\ref{fig:wtheta}.

We adopt the transition scale between the 2- and the 1-halo terms as
the radius of the proto-cluster, $r_{\rm cl}$. As illustrated by Fig.\,\ref{fig:wtheta}, the corresponding angular scale ranges from $\simeq 1^\prime$ to $\simeq 0.4^\prime$ for $z$ varying from $\simeq 1$ to $\simeq 5$. Hence, the \textit{Planck} resolution is not ideal for detecting individual proto-clusters. Higher resolution surveys, such as those planned with the Cosmic Origins Explorer (CORE), will be far more efficient for this purpose \citep{DeZotti2016}.

\begin{figure*}
\vspace{-5.0cm}
\hspace{-0.4cm}
\makebox[\textwidth][c]{
\includegraphics[width=1.0\textwidth]{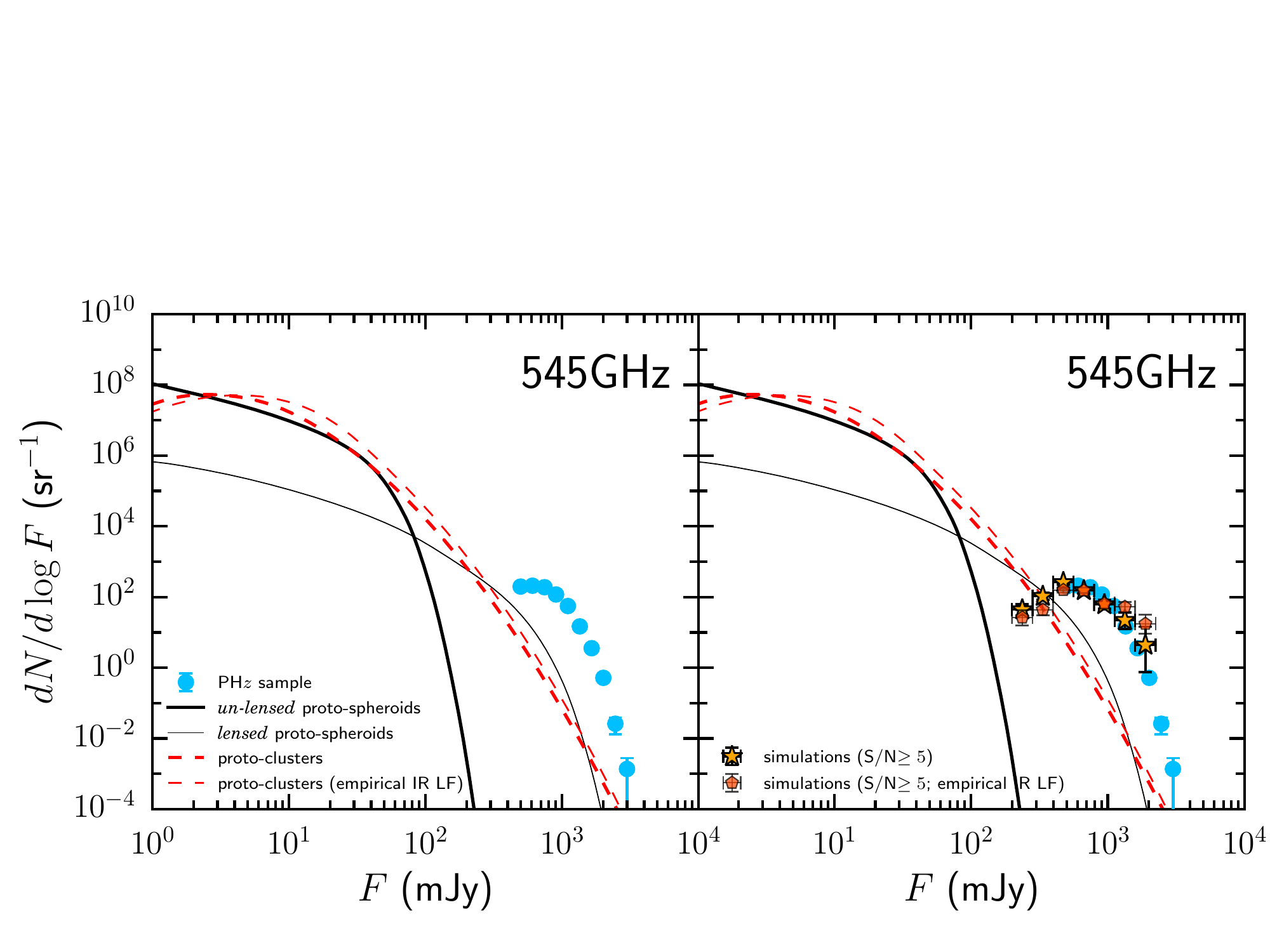}}
\vspace{-0.8cm}
\caption{Predicted differential number counts of proto-clusters at
  545\,GHz (550\,$\mu$m) derived from the Cai et al. model (2013; thick dashed red
  curve) and the infrared luminosity function of Mancuso et al. (2016;
  light dashed red curve). Data points are for the
  PH$z$ sample (PCXXXIX; blue circles). For comparison, the number counts of both
  un-lensed and strongly lensed proto-spheroidal galaxies, from the Cai et
  al. model, are
  shown by the thick and the thin solid black curves, respectively. On
  the right-hand panel the {\it Planck} results are compared with the number
  counts of sources detected with signal-to-noise ratio $SNR\geq5$ in the simulated map
  (yellow stars for the Cai model and orange pentagons for the
  empirical luminosity function of Mancuso et al.).}
\label{fig:dNdlgFclump_545GHz}
\end{figure*}

The additional power-law contribution to $w(\theta)$ dominates on scales $\simlt 0.1^\prime-0.2^\prime$ corresponding to \textit{physical} linear scales of 50--100 kpc. Interestingly, this is the scale of the proto-cluster cores at $z\simeq 2.4$--2.5 detected by \citet{Ivison2013} and \citet{Wang2016}.

The integrals over $r$ in eqs.\,(\ref{eq:meanL_clump}) and (\ref{eq:sigma2}) are carried out up to $r_{\rm sup}=r_{\rm cl}$, while the lowest value of $r$ is set by the minimum angular scale probed by the Chen et al. measurements, i.e. $\sim 0.025^{\prime}$ (or 1.5$^{\prime\prime}$), which corresponds to a \textit{comoving} length of $\sim25-64$\,kpc for $z=1-7$. In our calculations we set $r_{\rm inf}=40$\,kpc (comoving), corresponding to $\simeq 13\,$kpc (physical) at $z\simeq 2$ where the distribution of \textit{Planck} proto-cluster candidates peaks.
The values of
$r_{\rm inf}$ and $r_{\rm sup}$, as a function of redshift, are shown by the upper and lower limits
of the shaded region in lower panel of Fig.\,\ref{fig:scales}.
In the same figure, the dotted curve represents the
scale sampled by the beam of {\it Planck} at sub-millimeter
wavelengths (i.e. $5^{\prime}$). It is clear that such proto-clusters
would appear as point-like sources in the {\it Planck} maps.

For the amplitude $Q$ of the three point correlation function $\zeta$
we assumed $Q(z)=1/b^{2}(M_{\rm eff},z)$. This
formula was derived by \cite{Szapudi2001} from N-body simulations and
holds on scales $\lsim10\,$Mpc\,{\it h}$^{-1}$, i.e. on the scales of
interest here. We neglect any dependence of $Q$ on the
linear scale. At variance with N05 we do not consider a model with
$Q=1/b$ suggested by perturbation theory \citep{FG1993} as this holds
on larger scales, i.e. in the limit of small density
fluctuations. We also discard the unrealistic
model corresponding to $Q(z)=1$, as this holds for a pure dark-matter
distribution, i.e. not for visible galaxies
\citep[e.g.][]{Szapudi1996}. \\
We note that since $Q(z)$ decreases with increasing redshift, the
contribution to $\sigma^{2}_{\rm cl}$ from the second term in the
right-hand side
of eq.\,(\ref{eq:sigma2_clump}) may become negative. When this happens we set such contribution to zero.

\begin{table}
  \begin{center}
    \caption{Information about the proto-clusters at redshift $z>2$
      in Fig.\,\ref{fig:NgtLirclump}. The quoted infrared luminosities are those derived by summing
      the contributions from all the detected star forming galaxy
      members. Therefore they do not account for the emission due to
      unresolved/undetected cluster members.}\label{tab:info_protoclusters}
    \vspace{-0.0cm}
    \begin{tabular}{lll}
      \hline
      \hline
      $z$   & $L_{\rm IR}\,(\times10^{12}\,L_{\odot})$      &   Reference  \\
\hline
     2.05$\pm$0.09     &  31$\pm$6      &  \cite{Clements2014}  \\
      2.27$\pm$0.12    &  73$\pm$11    & \cite{Clements2014}  \\
     2.41                      &  57$\pm$6                  & \cite{Ivison2013}  \\
     2.506                    & 15.8             & \cite{Wang2016}  \\
     5.30                       &  17$\pm$8    & \cite{Capak2011}  \\
\hline
\hline
    \end{tabular}
  \end{center}
\end{table}

\section{Results and comparison with observations}\label{sec:results}

The search of proto clusters at high redshift have been carried out
using many different approaches and several of these structures have
already been found. We provide below a summary of some of the most
significant findings.

By exploiting the wealth of data covering the
entire accessible electromagnetic spectrum in the 2\,deg$^{2}$
Cosmological Evolution Survey \citep[COSMOS;][]{Scoville2007} field,
\cite{Capak2011}
searched for $z>4$ proto-clusters finding one at
$z=5.30$. This primordial cluster is characterized by an overdensity of
galaxies within  a region of $\sim2\,$Mpc
(comoving) radius around the extreme starburst galaxy
COSMOS AzTEC-3
and has an estimated total (baryonic+dark) mass of
$\gsim4\times10^{14}$\,M$_{\odot}$.

\citet{Ivison2013} identified at least
four intrinsically luminous galaxies at $z=2.41$ across an
$\sim100\,$kpc region by studying the CO line properties of the sub-mm
brightest sources in the
{\it Herschel} Astrophysical Terahertz Large Area Survey \citep[{\it
  H}-ATLAS;][]{Eales2010}. Specifically, they focused on a sample of
candidate lensed galaxies with $2.1\lsim z\lsim3.5$, extracted from
$\sim330\,$deg$^{2}$ of the {\it H}-ATLAS (i.e. the Equatorial $+$ North
Galactic pole fields), for which measurements of the CO(1-0) emission
line were previously obtained with
the Green Bank Telescope \citep{Harris2012}. The clump of luminous galaxies discovered by Ivison et al. was interpreted as
the core of an overdensity that will evolve into a massive
($\sim10^{14.6}\,$M$_{\odot}$) cluster.

\citet{Wang2016} looked instead for concentrations of $K_{s}$-band selected
red galaxies at $z>2$ over a 1.62\,deg$^{2}$ region of the COSMOS field, finding a significant overdensity of
massive galaxies at $z=2.506$ associated with extended X-ray
emission. The high star formation rate
($\sim3400\,$M$_{\odot}\,$yr$^{-1}$) measured in the
central 80\,kpc region of the overdensity suggests that also in this case
we are witnessing the rapid build-up of a cluster core.

\cite{Clements2014} examined the {\it Herschel}/SPIRE images of the 16
Planck Early Release Compact Source Catalog sources lying within the
90\,deg$^{2}$ of the Herschel Multitiered Extragalactic Survey \citep[HerMES;][]{Oliver12},
finding 4 overdensities of sub-mm sources. Using existing
multiwavelength photometric data they estimated redshifts in the
range $z\sim0.8-2.3$ and typical star formation rates SFR$\,>1000\,$M$_{\odot}$\,yr$^{-1}$ for
the protoclusters.

In Fig.\,\ref{fig:NgtLirclump} we compare these observations with the predictions based
on our model. The measured infrared luminosities of the
proto-clusters, listed in Table\,\ref{tab:info_protoclusters}, are
presented as lower limits. In fact they only account for cluster
members detected above the flux limit of the observations, while a significant, or even dominant, contribution is expected from fainter members. For example, using the \citet{Cai2013} model we find that sources brighter than 50\,mJy at $500\,\mu$m, approximately the 90\% completeness level of \textit{Herschel} observations used by \citet{PlanckCollaborationXXVII2015}, comprise 29, 43, 65 and 66\% of the luminosity obtained integrating the whole luminosity function at $z=1.5$, 2, 2.5 and 3, respectively. In other words, proto-clusters stand out more clearly in sub-mm maps, such as those that have been provided by \textit{Planck} and will be hopefully provided by CORE, than in much higher resolution point source surveys at the same wavelengths. This is in keeping with the findings by \citet{Clements2014} and \citet{PlanckCollaborationXXXIX2015} who noted that the sub-mm flux densities of proto-cluster candidates measured by \textit{Planck} are  about 2 to 3 times larger than the summed luminosities of member galaxies detected with \textit{Herschel} within the \textit{Planck} beam, although part of the difference is to be attributed to the `flux boosting' affecting the low signal-to-noise \textit{Planck} measurements (sources are preferentially detected if they are on top of positive confusion plus instrumental noise fluctuations).

Not only the proto-cluster luminosities but also the number densities inferred from the \citet{Ivison2013} and \citet{Wang2016} detections are lower limits. \citet{Ivison2013} only focused on {\it Herschel} sources with existing CO line measurements, while \citet{Wang2016} based their selection on a $K_{s}$-band catalogue, that may miss proto-clusters of heavily dust-enshrouded galaxies. Information on the redshift and infrared luminosity of the proto-clusters in Fig.\,\ref{fig:NgtLirclump} are summarized in Table\,\ref{tab:info_protoclusters}.

\begin{figure}
\vspace{0.0cm}
\hspace{-4.4cm}
\makebox[\textwidth][c]{
\includegraphics[width=0.54\textwidth]{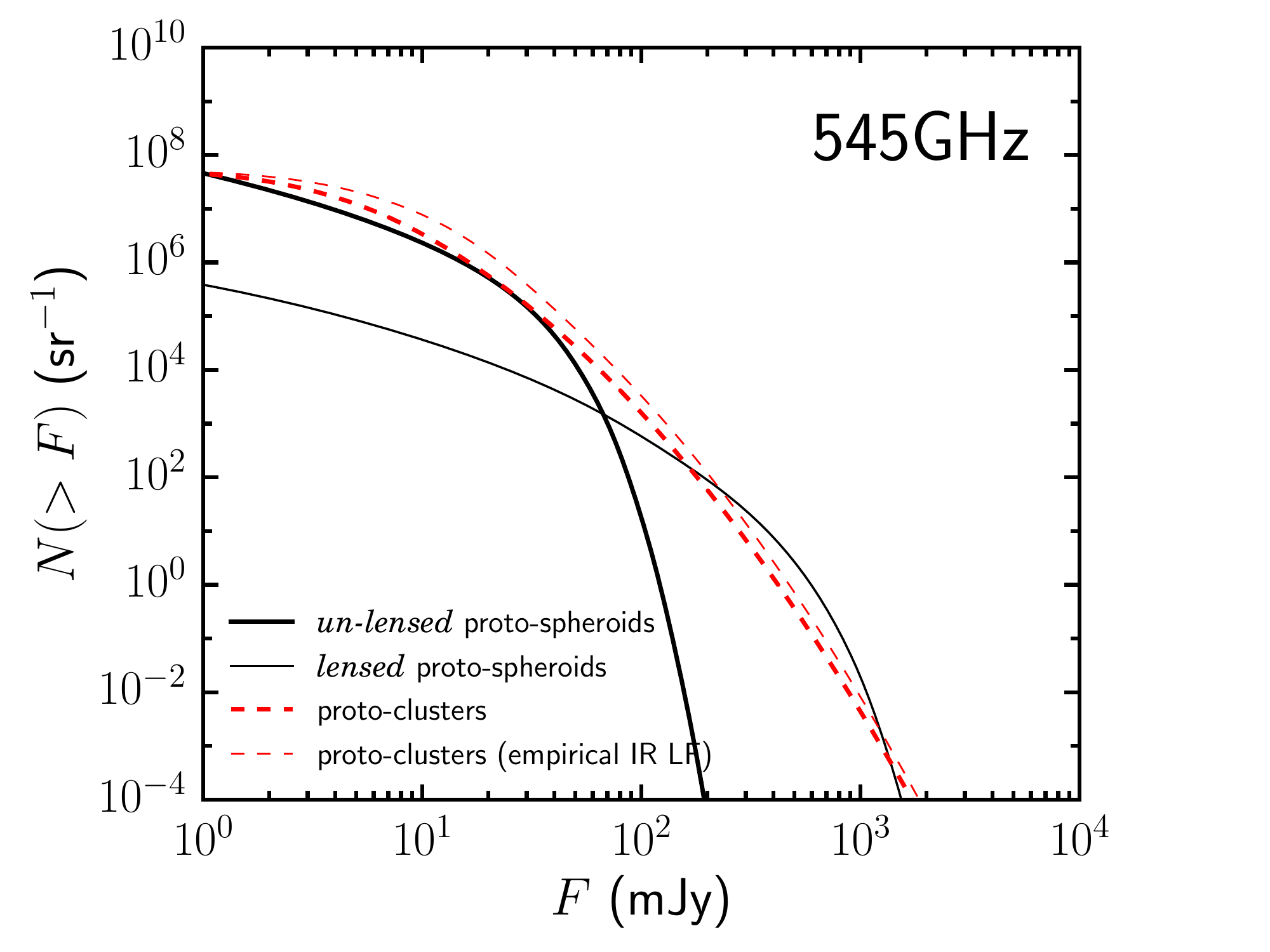}}
\vspace{-0.6cm}
\caption{Predicted integral number counts of proto-clusters at
  545\,GHz (550\,$\mu$m; dashed red curves). The meaning of the curves
  is the same as in Fig.\,\ref{fig:dNdlgFclump_545GHz}.}
\label{fig:NgtFclump_545GHz}
\end{figure}

We also show, in the same figure, the predictions based on the
  infrared luminosity function of
  \cite{Mancuso2016}. For $2\lsim z\lsim3.5$ the latter predicts
  $\times1.5-2$ more proto-clusters than the Cai et
  al. model. This is because the Mancuso et al. infrared luminosity
  function
is higher than the Cai et al. one for $L_{\rm
    IR}\lsim10^{12}\,$L$_{\odot}$ and $z\gsim2$, which leads to
  higher values of the mean and variance of the
  clump luminosity. However, the
  difference between the two models is still consistent with the uncertainties in the
  measured luminosity function of dusty galaxies at those luminosities
  and redshifts \citep[see fig.1 of][]{Mancuso2016}.

Having checked that our model yields results consistent with
observations, that refer to the redshift range probed by the sample of
PCXXXIX, we compare, in Fig.~\ref{fig:dNdlgFclump_545GHz}, our
predictions (dashed red curve; thicker for the Cai et al. model
  and lighter for the Mancuso et al. empirical luminosity function) with the counts at 545\,GHz
(550$\,\mu$m) of the PH$z$ sources, reported in the latter paper
(filled blue circles).  The number counts at a given wavelength are
derived from the IR luminosity function of the clumps
[eq.\,(\ref{eq:LF_clumps})] by adopting the spectral energy
distribution (SED) of the $z=2.3$ star forming galaxy SMM\,J2135-0102
\citep{Swinbank2010} for the proto-spheroids, as done by
\cite{Cai2013}.

Apart from the small fraction of strongly lensed
galaxies \citep[around 3\%;][]{PlanckCollaborationXXVII2015}, these
candidate high-$z$ sources may be either over-densities of bright
star-forming galaxies (i.e. proto-clusters) or high peaks of the
cosmic infrared background (CIB) fluctuations with red colours because
are contributed by physically unrelated high-$z$ galaxies.

Figure\,\ref{fig:dNdlgFclump_545GHz} also shows, for comparison, the differential number
counts of strongly lensed galaxies.
Such counts were computed using the
formalism of \citet[][their SISSA model]{Lapi2012} with a maximum
amplification $\mu_{\rm max}=15$ that is found by \cite{Neg2016}
to reproduce the counts of strongly lensed galaxies in the {\it
  H}-ATLAS. The number densities of the PH$z$ sample are well in excess of those expected from
our model, also when using the empirical luminosity function of
  Mancuso et al., and, more importantly, of the number densities of
high-$z$ halos capable of hosting them,
as pointed out in Sect.~\ref{sec:intro}. In the next Section we discuss a plausible explanation of the discrepancy.

\begin{figure*}
  \hspace{-1.0cm}
  \begin{minipage}[b]{0.51\linewidth}
    \centering \resizebox{1.0\hsize}{!}{
      \hspace{+1.3cm}\includegraphics{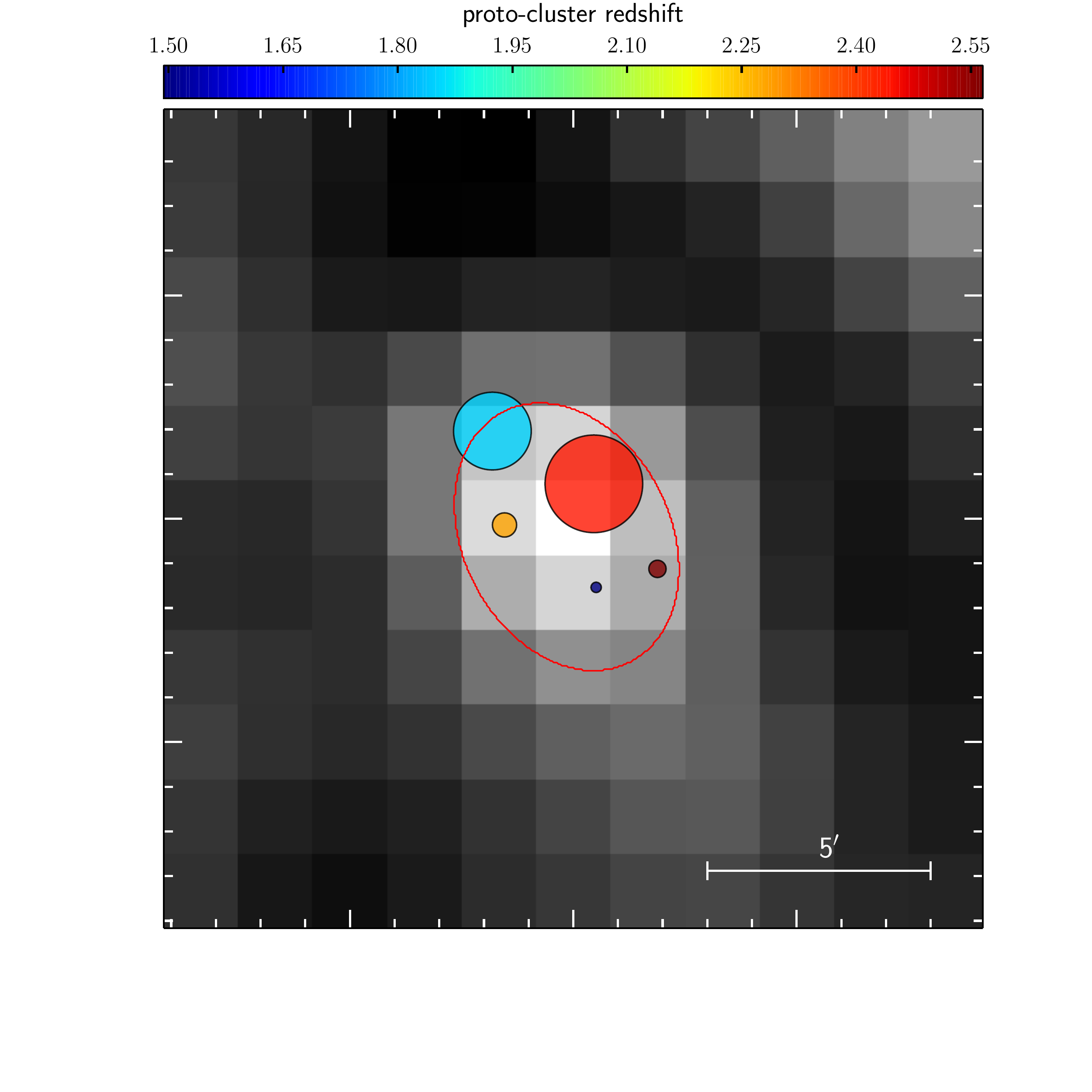}} 
  \end{minipage}
  \hspace{0.0cm}
  \begin{minipage}[b]{0.51\linewidth}
    \centering \resizebox{1.0\hsize}{!}{
      \hspace{+1.3cm}\includegraphics{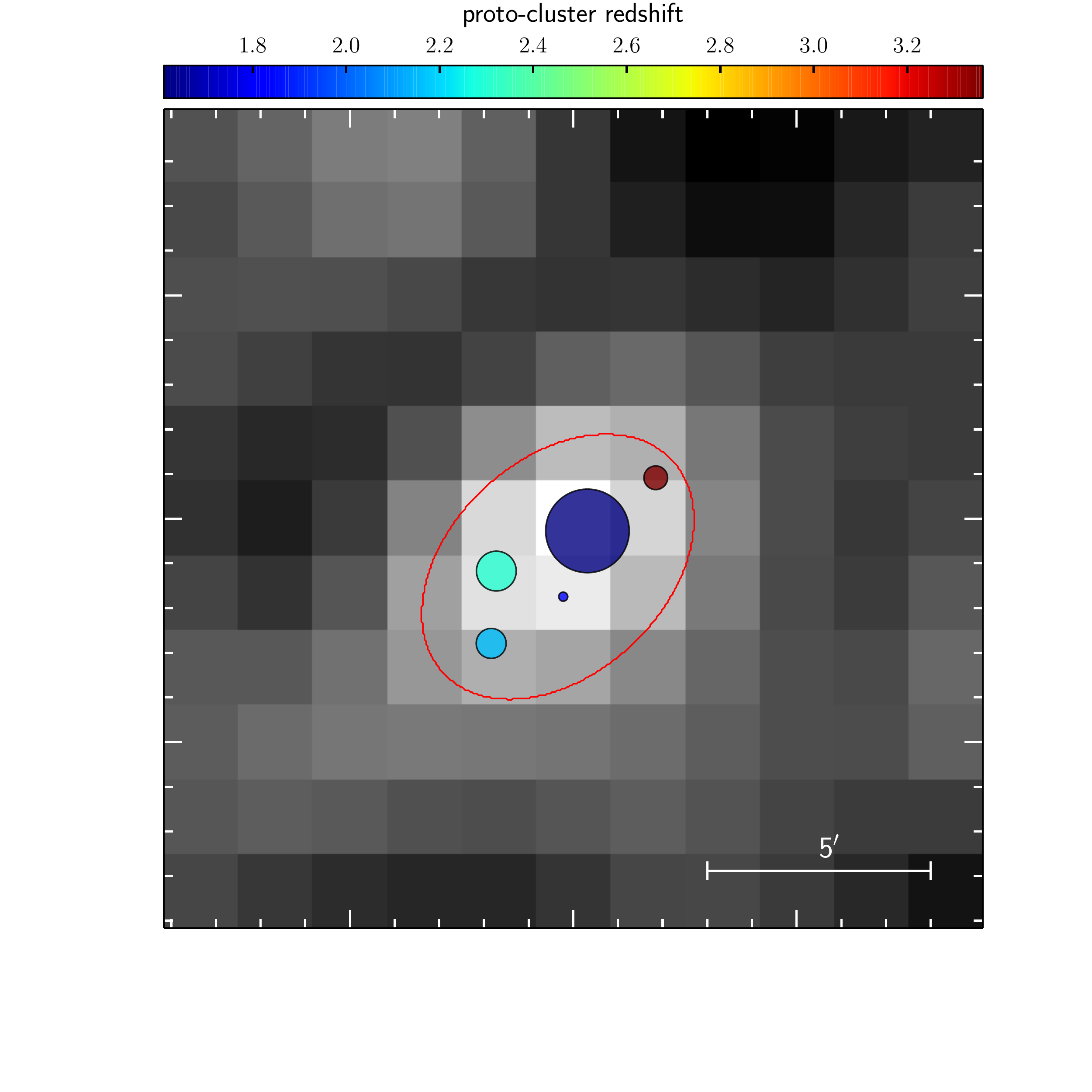}} 
  \end{minipage}
 \vspace{-1.1cm}
 \caption{Examples of sources detected in the 545\,GHz simulated map
   (black and white image in the background). The red ellipsis is the
   Gaussian fit to the detection,
   while  the spots mark the five brightest proto-clusters within the ellipsis,
   scaled in size according to their flux  density (i.e. the brighter the clump, the bigger the spot).}
 \label{fig:detections_examples}
\end{figure*}

\section{The effect of source confusion}\label{sec:simulations}

Thanks to the very low instrumental noise of the \textit{Planck} High
Frequency Instrument (HFI), the fluctuation field measured in the HFI maps
is signal-dominated. In regions with low Galactic dust content,
clustering of high-$z$ sources making up the CIB is the dominant
source of intensity fluctuations, at sub-mm wavelengths, on the
several arcmin scales of interest here. The probability distribution
function (PDF) of such intensity fluctuations is highly skewed, with
an extended tail towards high signal-to-noise ratios
\citep[cf. Fig.~10 of][]{DeZotti2015}.

To investigate the nature of high signal-to-noise intensity peaks we
have resorted to Monte Carlo simulations. Proto-clusters
have been randomly distributed
in infrared luminosity and redshift over an area of $5000$\,deg$^{2}$,
according to the modelled luminosity functions. The area was divided into
1.7$^{\prime}$$\times$1.7$^{\prime}$ pixels\footnote{1.7$^{\prime}$
  corresponds to 1/3 of the {\it Planck} Full Width at Half Maximum
  (FWHM) at 545\,GHz.} and the flux densities of proto-clusters
within each pixel were summed up. The map was then convolved with the
{\it Planck} beam
and background
subtracted.
The source extraction was performed by looking
for connected pixels with signal-to-noise ratio (SNR)\footnote{A detection requires at least one pixel above the
  SNR threshold.} SNR$\ge5$ as in PCXXXIX. A couple of examples of the
content of simulated intensity peaks with $\hbox{SNR}\ge 5$ are shown
in Fig.~\ref{fig:detections_examples}. One or two very luminous
proto-clusters are generally present; however, most of the flux
density within the \textit{Planck} beam is accounted for by much
fainter objects, as implied by the steepness of the bright portion of
the proto-cluster counts (Figs.~\ref{fig:dNdlgFclump_545GHz} and
\ref{fig:NgtFclump_545GHz}).

The source flux density is obtained by first fitting a Gaussian to a
$20^{\prime}\times 20^\prime$ postage stamp centered on the detected
object and then integrating the best-fit model.
The differential number counts measured in
the simulated map are shown in the right-hand panel of
Fig.\,\ref{fig:dNdlgFclump_545GHz} by the yellow stars. The effect of source confusion
is dramatic and brings the number counts in line with the
measured ones (blue dots). The same conclusion is reached when
  we use the Mancuso et al. empirical luminosity function for dusty
  galaxies, as illustrated by the orange pentagons in the same figure.

The model integral number counts of \textit{individual} high-$z$ ($z\simgt 2$) proto-clusters, displayed in Fig.~\ref{fig:NgtFclump_545GHz}, show that those brighter than the flux limit of the PH$z$ sample (500\,mJy at 545 GHz), hence detectable by \textit{Planck}, are expected to be very rare. The CORE project can do much better. Its estimated $4\,\sigma$ detection limits, at the nearby frequency of 520\,GHz, range from 141\,mJy for the 1\,m telescope option to 82.5\,mJy for a 1.5\,m telescope \citep{DeZotti2016}. From Fig.~\ref{fig:NgtFclump_545GHz} the corresponding surface densities range from $230\,\hbox{sr}^{-1}$ to $3300\,\hbox{sr}^{-1}$.

\begin{figure}
\vspace{-0.75cm}
\hspace{-4.4cm}
\makebox[\textwidth][c]{
\includegraphics[width=0.55\textwidth]{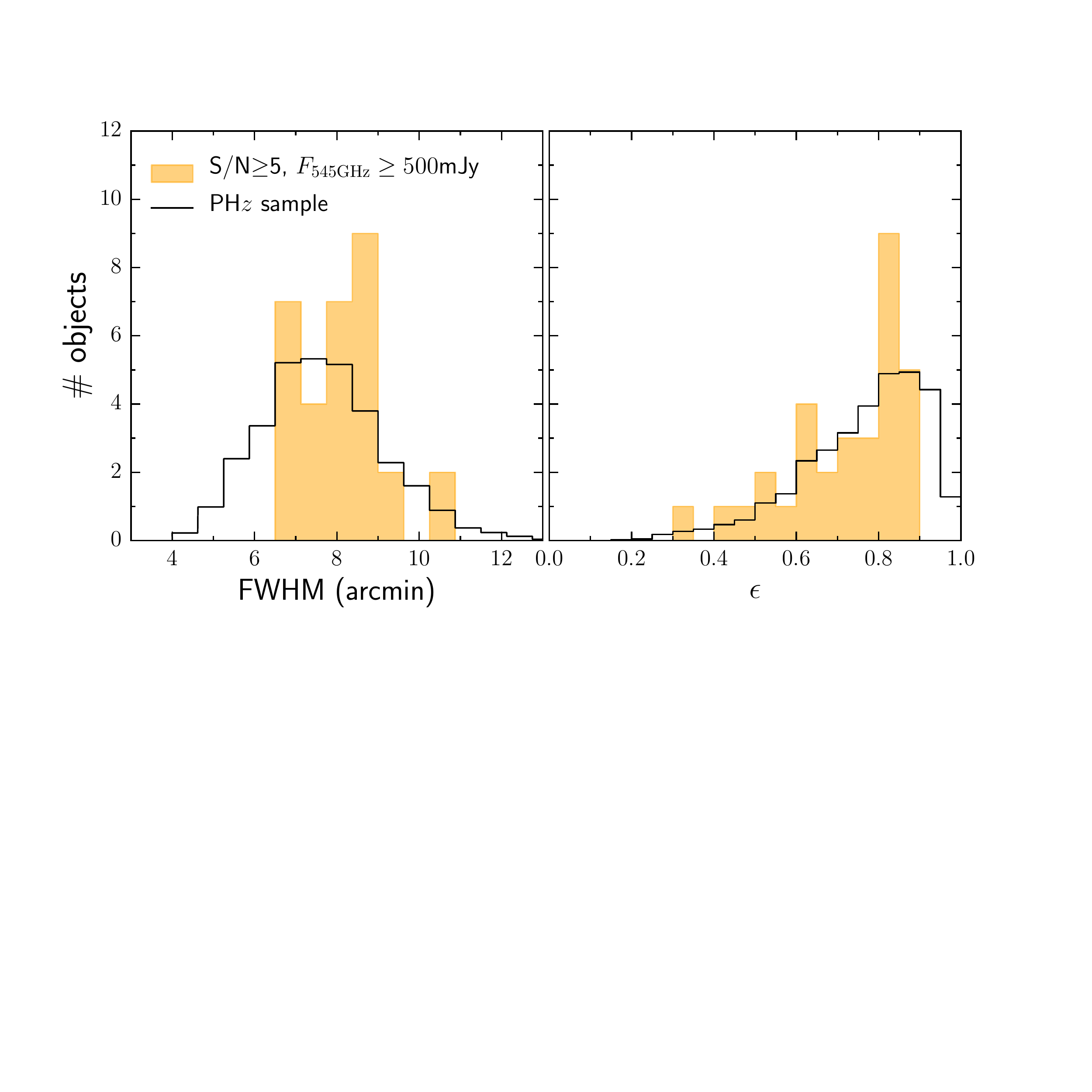}}
\vspace{-4.8cm}
\caption{Distribution of FWHMs and ellipticities,
  $\epsilon$, derived from the simulations for detections with ${\rm
    SNR}\ge5$ and $F_{\rm 545GHz}\ge500\,$mJy (yellow histograms),
  compared to those measured for the PH$z$ sample, normalized to the number of sources in the simulation (solid line). The FWHM is defined as $({\rm FWHM}_{\rm major}\times
  {\rm FWHM}_{\rm minor})^{1/2}$ where ${\rm FWHM}_{\rm major}$ and ${\rm FWHM}_{\rm minor}$ are the major
and the minor FWHM derived from the Gaussin fit, while the ellipticity is
defined as $\epsilon=[1 - ({\rm FWHM}_{\rm minor}/{\rm FWHM}_{\rm major})^{2}]^{1/2}$. Note that the simulations cannot reproduce accurately the selection criteria of the PH$z$ sample, nor the procedures for measuring the FWHM and the ellipticity. Hence a precise match between the model and the observed distributions cannot be expected (see text). }
\label{fig:FWHM_epsilon}
\end{figure}

The yellow histograms in Fig.\,\ref{fig:FWHM_epsilon} show the distributions of the FWHMs and of the ellipticities, $\epsilon$, of
sources detected in simulated maps with SNR$\geq5$  and $F_{\rm
  545GHz}\ge500\,$mJy (very similar results are obtained using
  the  empirical luminosity function of Mancuso et al.). The distributions are qualitatively consistent with those of the PH$z$ sample (solid histograms, taken from Fig.\,14 of PCXXXIX), suggesting that the elongated shape of the detected sources is an effect of source confusion rather than being an intrinsic property of the proto-clusters. Note that a close agreement between the results of our simulations and the observed distributions is not to be expected because simulations cannot reproduce exactly the selection criteria adopted by PCXXXIX (simultaneous detection within a $5^\prime$ radius in the 545 GHz excess map, with $\hbox{SNR}>5$, and in the 857, 545, and 353 GHz cleaned maps with $\hbox{SNR}>3$; absence, at 100\,GHz, of any local maximum with $\hbox{SNR}>3$  within a radius of $5^\prime$ of the 545\,GHz position; colour-colour selection) and their procedures for determining sizes and ellipticities.

\begin{figure}
\vspace{0.0cm}
\hspace{-4.4cm}
\makebox[\textwidth][c]{
\includegraphics[width=0.54\textwidth]{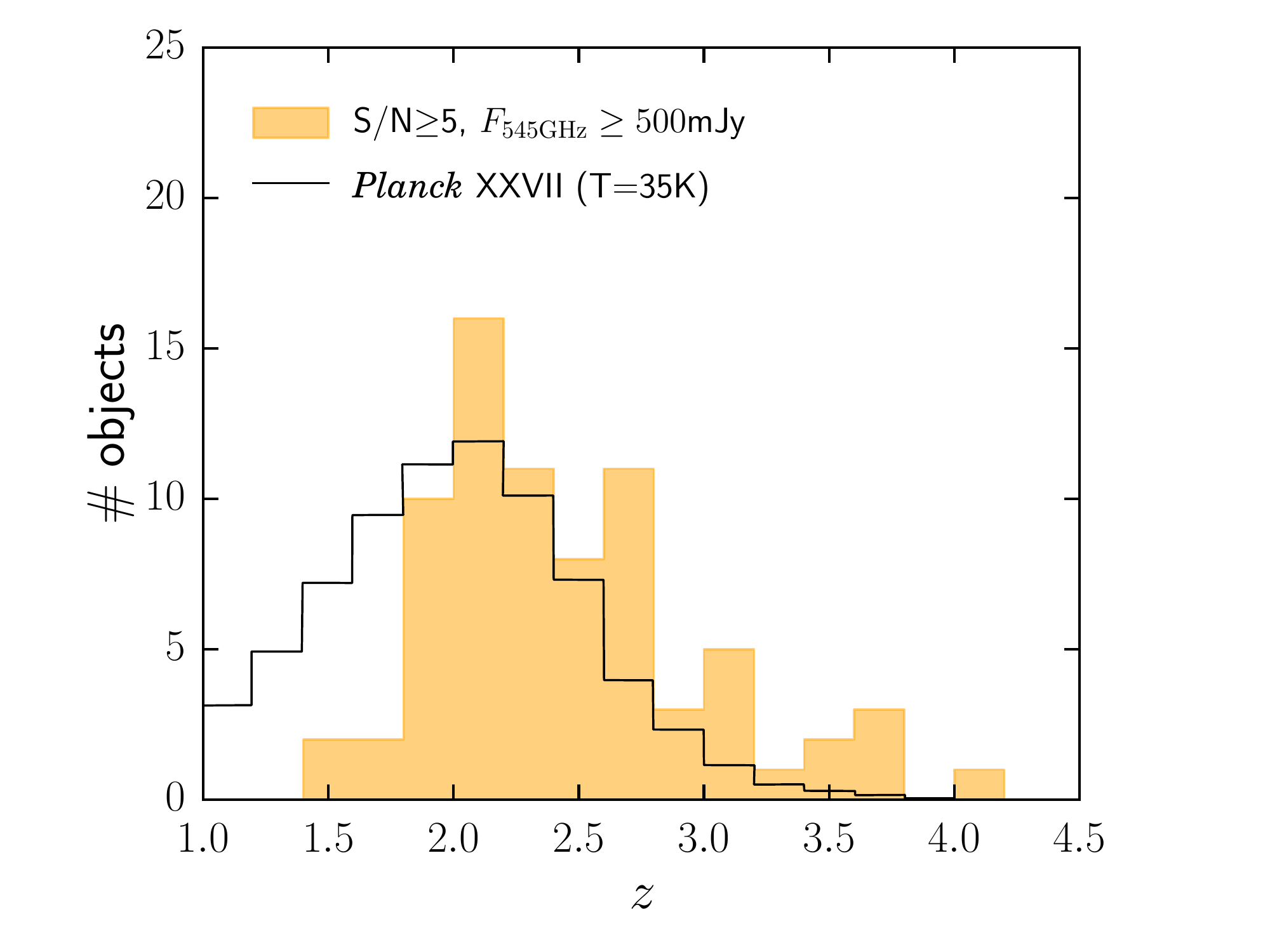}}
\vspace{-0.6cm}
\caption{Redshift distribution of the brightest proto-clusters
  associated with the sources detected in the simulation with ${\rm
    SNR}\ge5$ and $F_{\rm 545GHz}\ge500\,$mJy (yellow histogram). For
  comparison we show the photometric redshift distribution of the {\it
    Planck} candidate proto-clusters (Planck Collaboration XXVII 2015) derived from
   the associated {\it Herschel}/SPIRE sources (black line) assuming a modified black body spectrum with dust temperature
$T=35\,$K and dust emissivity index $\beta=1.5$.}
\label{fig:Nz_clumps}
\end{figure}

Finally, in Fig.\,\ref{fig:Nz_clumps} we show the redshift distribution of the brightest proto-clusters associated with the
sources detected in the simulated map with SNR$\geq5$ and $F_{\rm
  545GHz}\ge500\,$mJy. For comparison, we also show the
photometric redshift distribution (obtained
assuming a modified black body spectrum with dust temperature
$T=35\,$K and dust emissivity index $\beta=1.5$) of the {\it
  Herschel}/SPIRE detected sources
associated with the {\it Planck} candidate proto-clusters \citep[][
black histogram in the same figure]{PlanckCollaborationXXVII2015}.
Both histograms peak in the redshift range $z=1.5-3$ showing
the great potential offered by the {\it Planck} catalogue for studying the
early phases of clusters formation.

\section{Conclusions}\label{sec:conclusions}

The PHz catalogue appears to be dominated by
over-densities of star-forming galaxies plus a small fraction of
strongly gravitationally lensed galaxies \citep[around
3\%;][]{PlanckCollaborationXXVII2015}. Motivated by these results, we
have updated the N05  predictions for the number
counts of proto-clusters of star-forming galaxies and compared them
with observational estimates.

We stress that our results are largely model-independent. The basic
ingredients of our calculations are the sub-mm luminosity functions of
high-$z$ galaxies and their spatial correlation function. The
luminosity functions have been observationally
determined up to $z\simeq 4$, based on  \textit{Herschel} survey data
\citep{Gruppioni2013}.
The model we have used reproduces very well the observational
determinations at each redshift
\citep{Cai2013,Bonato2014}. Furthermore, we have shown that the results do not change significantly
if the model is replaced by an empirical fit of the  observed
luminosity function of dusty galaxies \citep{Mancuso2016}. As shown by
Fig.~\ref{fig:wtheta}, our model also provides an accurate representation of the
observed redshift dependent correlation functions.

We find that the counts of sub-mm bright proto-clusters, obtained in the
framework of the standard
$\Lambda$CDM cosmology, are well below the observational estimates by
PCXXXIX. A similar conclusion can be inferred from the results of \cite{Granato2015}
who also predict, by using a completely different approach, a number of
proto-clusters well below the one suggested by the statistics of {\it Planck} detections.

We have shown that this
basically follows from the fact that, at high $z$, there are not
enough sufficiently massive halos. On the other hand we have shown, by means of simulations, that the
\textit{Planck} {\it cold} intensity peaks are fully consistent with
being mostly random fluctuations in the number of unrelated
proto-clusters at $z\ge 2$ within
the \textit{Planck} beam. This is highlighted in Fig.~\ref{fig:detections_examples} where
examples of sources detected in the
simulation are shown together with the brightest
proto-clusters they are comprised of. The counts of these fluctuations match very
well those of {\it cold}
peaks and there is qualitative agreement also with the distributions of their FWHM’s and of their ellipticities.
Interestingly, the only \textit{Planck} over-density for which spectroscopic or
photometric redshifts
of member galaxies have been obtained, was found to consist of 2
physically unrelated structures at $z\sim1.7$ and $z\sim2$
\citep{FloresCacho2015}.

The redshift distribution of the brightest proto-clusters contributing
to the {\it cold} peaks, given by the model, has a broad maximum
between $z=1.5$ and $z=3$.
Therefore, follow-up observations of galaxies within the Planck
overdensities providing redshift estimates would
be a powerful tool to investigate the early phases of cluster formation, inaccessible by other means.

$~$ \\
{\bf Acknowledgments} \\
We thank the referee for helpful suggestions.
MN acknowledges financial support from the European Union's Horizon
2020 research and innovation 
programme under the Marie Sk{\l}odowska-Curie grant agreement No 707601.
GDZ acknowledges financial support by by ASI/INAF agreement
n. 2014-024-R.1 for the {\it Planck} LFI activity of Phase
E2. J.G.N. acknowledges financial support from the Spanish MINECO for
a 'Ramon y Cajal' fellowship (RYC-2013-13256) 
and the I+D 2015 project AYA2015-65887-P (MINECO/FEDER)

\end{document}